\documentclass[jkps,twocolumn,fleqn,showpacs,showkeys]{revtex4}

\usepackage{graphicx}
\usepackage{amssymb}
\usepackage{amsmath}
\usepackage{bm}
\usepackage{times}

\newcommand{\msun}{M_\odot}
\newcommand{\Mbh}{M_{\rm BH}}

\newcommand{\chibh}{\chi_{\rm BH}}
\newcommand{\Mns}{M_{\rm NS}}
\newcommand{\be}{\begin{equation}}
\newcommand{\ee}{\end{equation}}
\newcommand{\bea}{\begin{eqnarray}}
\newcommand{\eea}{\end{eqnarray}}
\newcommand\flow{f_{\rm low}}

\newcommand{\etal}{{\it et al.}}

\begin{document}
\setcounter{page}{1}

\title[]{Accuracy in Measuring the Neutron Star Mass in Gravitational Wave Parameter Estimation for Black Hole-Neutron Star Binaries}

\author{Hee-Suk \surname{Cho}}
\email{chohs1439@kisti.re.kr}

\affiliation{Korea Institute of Science and Technology Information, Daejeon 305-806, Korea}

\date[]{}

\begin{abstract}
Recently, two gravitational wave (GW) signals, named as GW150914 and GW151226, have been detected by the two LIGO detectors.
Although both signals were identified as originating from merging black hole (BH) binaries, GWs from systems containing neutron stars (NSs) are also expected to be detected
in the near future by the Advanced detector network.
In this work, we assess the accuracy in measuring the NS mass ($\Mns$) for the GWs from BH-NS binaries adopting the Advanced LIGO sensitivity with a signal-to-noise ratio of 10.
By using the Fisher matrix method, we calculate the measurement errors ($\sigma$) in $\Mns$ 
assuming the NS mass of $1 \leq \Mns/\msun \leq 2$ and low mass BHs with the range of $4 \leq \Mbh/\msun \leq 10$.
We used the TaylorF2 waveform model where the spins are aligned with the orbital angular momentum, but here we only consider the BH spins.
We find that the fractional errors ($\sigma/\Mns \times 100$) are in the range of $10\% - 50\%$ in our mass region for a given dimensionless BH spin as $\chibh = 0$.
The errors tend to increase as the BH spin increases, and this tendency is stronger for higher NS masses (or higher total masses).
In particular, for the highest mass NSs ($\Mns=2~\msun$), the errors $\sigma$ can be larger than the true value of $\Mns$ if the dimensionless BH spin exceeds $\sim 0.6$.  

\end{abstract}

\pacs{04.30.–w, 04.80.Nn, 95.55.Ym}

\keywords{Gravitational waves, Parameter estimation, Fisher matrix, Black hole, Neutron star}

\maketitle

\section{INTRODUCTION}
During the first observational run of the Advanced LIGO detectors, the first gravitational wave signals were detected,
and further detailed analyses showed that these GWs were emitted from merging binary black holes  (BBHs) \cite{GW1,GW1PE1,GW2,GW1PE2}.
These observations indicate that  future observing runs of the Advanced detector network will yield many more BBH merger signals \cite{Aba10,Dom14,Abb16}.
On the other hand, it is expected that GW signals from black hole (BH)-neutron star (NS) or NS-NS binaries will also be captured a few times per year  in the near future \cite{Aba10,Dom14,Kim15},
although those signals will not be as many as the BBH signals.

Detection of GWs from BH-NS or NS-NS binaries is very important because those signals can tell us about the nature of NSs, particularly the NS masses.
Determining the upper limit of the NS mass is one of the most challenging issues in modern astrophysics.
In various  theoretical models, the NS masses are expected to range between 1 $\msun$ and 3 $\msun$ \cite{Lat07}.
On the other hand,  in most of the well-measured NS-NS or NS-white dwarf binaries, the NS masses seem to cluster around $\sim 1.4 ~ \msun$~\cite{Pra13} except the two higher mass NSs whose masses are $\sim 2 ~ \msun$~\cite{Dam10,Ant13}.
However, more observations are
still necessary to robustly confirm the NS mass limit,
and GWs from BH-NS or NS-NS binaries would provide a tighter constraint on the distribution of NS masses.

In GW data analysis for compact binary coalescences, once a detection is made in the search pipeline,
a detailed analysis is performed in the parameter estimation pipeline to identify the physical parameters of the binary source \cite{Aas13,Cho15a,GW1PE1}.
The result of the parameter estimation is  given by a probability density distribution in the parameter space.
Typically, a very long computation time is required to complete the parameter estimation procedure.  
However, the measurement errors can be easily approximated by using the Fisher matrix (FM) method if the signal is strong enough and the noise is Gaussian.

In the previous work \cite{Cho15b}, we showed the measurement accuracy of the NS mass with various companion masses 
for a nonspinning binary system. In this work, we extend the previous work to a more generic system where the binaries can have spins.
Particularly, we only consider the BH spin because the contribution of the NS spin to the binary evolution is negligible compared to that of the BH spin.
Furthermore, we also assume that the BH spin angular momentum is aligned with the orbital angular momentum so that the binary does not precess during the evolution.
As in the previous work, we adopt a simple Fourier-domain waveform model and use the FM method to 
predict the measurement errors of the NS mass with the Advanced LIGO detector sensitivity.



\section{waveform model for aligned-spin binaries}
In the past studies on GW data analysis, the most commonly used waveform model is the TaylorF2, which is a Fourier-domain 
model obtained from a time-domain post-Newtonian model via the stationary phase approximation \cite{Sat91,Cut94,Poi95}. 
The waveform function of the TaylorF2 is expressed  as
\be \label{eq.TaylorF2}
h(f)=Af^{-7/6} e^{i \Psi(f)},
\ee
where $A$ is the wave amplitude that consists of the binary masses and the extrinsic parameters.
The amplitude simply sets a scale for the matched filter output, so does not affect our analysis.
The wave phase is defined as
\begin{equation} \label{eq.Psi}
\Psi(f)=2\pi ft_c -2 \phi_c -{\pi\over 4} + {3 \over 128 \eta v^5} \phi(f),
\end{equation}
where $t_c$ and $\phi_c$ are the coalescence time and the phase at coalescence instant, $\eta=m_1m_2/M^2$ is the symmetric mass ratio with $M=m_1+m_2$
and $\phi(f)$ can be expressed by using the post-Newtonian expansion as~\cite{Aru09}
\begin{widetext}
\bea
\phi(f) & = &1 
+\left( \frac{3715}{756} + \frac{55}{9}\eta \right)v^2 
+(4\beta- 16\pi) v^3 
+ \left( \frac{15293365}{508032} + \frac{27145}{504}\eta + \frac{3085}{72}\eta^2 -10 \sigma  \right) v^4 \\ 
&+& \left(\frac{38645 \pi}{756} - \frac{65 \pi }{9}\eta  -\gamma \right)
(1  + 3\log v) v^5
+ \bigg\{ \frac{11583231236531}{4694215680} - \frac{640}{3}\pi^2 -
\frac{6848\gamma_{\rm E} }{21} -\frac{6848 \log\left(4{v}\right)}{ 21}   \nonumber\\ 
&+&  \left( 
 \frac{2255{\pi }^2}{12}  - \frac{15737765635}{3048192}\right)\eta 
+\frac{76055}{ 1728}\eta^2-\frac{127825}{ 1296}\eta^3
\bigg\} v^6 
+ \left(\frac{77096675 \pi}{254016} + \frac{378515\pi}{1512}\eta 
- \frac{74045\pi}{756}\eta^2\right)v^7,\nonumber
\label{eq.SPA}
\eea
\end{widetext}
where $v= [\pi f M]^{1/3}$,  $\gamma_{\rm E}=0.577216...$ is the Euler constant, and 
the terms $\beta, \sigma,$ and $\gamma$ denote the leading-order spin-orbit coupling, leading-order spin-spin coupling, and next-to-leading-order spin-orbit
coupling, respectively. For an aligned-spin system, these can be expressed as
\begin{widetext}
\bea
\beta &=& \sum^2_{i=1} \left[\frac{113}{12}\left( \frac{m_i}{M}\right)^2+\frac{25\eta}{4}\right]\chi_i,  \ \ \ \ \ \ \
\sigma=\frac{474\eta}{48} \chi_1 \chi_2 +\sum_{i=1}^2\frac{81}{16} \left( \frac{m_i}{M} \right)^2 \chi_i^2\,   \nonumber\\
\gamma&=&\sum_{i=1}^2\left[ \left( \frac{732 985}{2268}+\frac{140\eta}{9}  \right)  \left( \frac{m_i}{M} \right)^2+ \eta \left(  \frac{13915}{84}-\frac{10\eta}{3}  \right) \chi_i \right]
\label{eq.SPA}
\eea
\end{widetext}
where   $\chi_i \equiv S_i/m_i^2$ is a dimensionless BH spin, $S_i$ being the spin angular momentum of the $i{\rm th}$ compact object. 
For a nonspinning system, the above spin terms are simply set to be 0, and 
we only need to consider four physical parameters; $\{m_1, m_2, t_c, \phi_c\}$.
On the other hand, for an aligned-spin system the two parameters $\chi_1$ and $\chi_2$ should also be included.
However, since we decided to only consider  the BH spin in this work, we remove the second spin component by choosing as $\chi_2\equiv \chi_{\rm NS}=0$.
Thus, we will deal with the five parameters; $\{m_1, m_2, \chi_1, t_c, \phi_c\}$ in our analysis.


\section{Measurement error in Parameter Estimation}
A match between a detector data stream $x(t)$ and a model waveform $h(t)$ can be obtained by using the inner product $\langle ... | ... \rangle$ as
\be \label{eq.match}
\langle x | h \rangle =  4 {\rm Re} \int_{\flow}^{\infty}  \frac{\tilde{x}(f)\tilde{h}^*(f)}{S_n(f)} df,
\ee
where $\tilde{x}(f)$ represents a Fourier transform of $x(t)$,
$S_n(f)$ is the noise power spectral density for the detector
and $\flow$ is the low-frequency cutoff that depends on the shape of $S_n(f)$. 
We take into account the zero-detuned, high-power  noise power spectral density of Advanced LIGO \cite{psd},
and $\flow$ is chosen to be 10 Hz.

The parameter estimation algorithm performs the above match computations iteratively until the algorithm  recovers the true values of the parameters.
Thus, the efficiency of parameter estimation is subject to how fast the algorithm can find the true values, and
its computation time mainly depends on the waveform model and a dimension of the parameter space. 
Generally, this procedure is a time-consuming  task because the parameter estimation explores the whole parameter space without any information about the true parameters except the coalescence time \cite{GW1PE1,GW1PE2}.
On the other hand, if an incident GW signal is buried in Gaussian noise and strong enough,  
the posterior probability density function used for parameter estimation is given by a Gaussian distribution of the form \cite{Fin92}
\begin{equation} \label{eq.p}
p(\Delta \lambda^i) \propto {\rm exp} [-\frac{1}{2}\Gamma_{ij}\Delta \lambda^i \Delta \lambda^j],
\end{equation}
where $\Gamma_{ij}$ is the FM defined as~\cite{Val08,Jar94,Cho13}
\be\label{eq.FM}
\Gamma_{ij}=-\rho^2 \bigg \langle {\partial \hat{h} \over \partial \lambda_i} \bigg| {\partial \hat{h} \over \partial \lambda_j} \bigg \rangle\bigg|_{\lambda=\lambda_{\rm true}},
\ee
where $\rho\simeq \langle h | h\rangle$ is the signal-to-noise ratio (SNR) and $\hat{h}\equiv h/\sqrt{\langle h | h\rangle}$ is the normalized waveform.
The inverse of $\Gamma_{ij}$ represents the covariance matrix of the parameter errors, 
and the error ($\sigma_i$) of each parameter is determined by $\sigma_i=\sqrt{(\Gamma^{-1})_{ii}}$.
Thus,  $\sigma_i$ is inversely proportional to the SNR.

For our BH-NS system, by applying the phase function in Eq. (\ref{eq.Psi}) to the FM formalism in Eq. (\ref{eq.FM}), 
we can calculate a $5 \times 5$ matrix for the parameters $\{\Mbh, \Mns, \chibh, t_c, \phi_c\}$ with true values given,
and obtain the measurement error $\sigma_i$ of each parameter.
We do not present the errors of $t_c$ and $\phi_c$ in our results because these two parameters are arbitrary constants.
However, unlike the extrinsic parameters incorporated in the wave amplitude, these are in general strongly correlated with the other parameters.
Therefore, they should be taken into account simultaneously with the other parameters in the construction of the FM.

In order for the FM approach to be valid, two conditions are required: Gaussian noise and high SNR.
If non-Gaussian noises are mixed in the data, the posterior probability density function can show a non-Gaussian distribution and the recovered 
parameter values can be biased. However, such non-Gaussian noises are unlikely to make a significant impact on the parameter  estimation result 
if the SNR is high enough \cite{Ber15}. In this work, we assume a moderately high SNR of 10.
On the other hand, it has been noted that due to the abrupt cutoff of the waveforms at the innermost stable circular orbit ($\pi M f_{\rm isco}=6^{-3/2}$), the FM result obtained by using the TaylorF2 model cannot be trusted in the high mass region beyond $\sim 10~ \msun$ \cite{Rod13,Man14,Cho14}.
Therefore, we only consider the BH mass up to $10 ~\msun$.


\section{Result: Measurement error for the neutron star mass}
Here, we  present our result of measurement errors in $\Mns$ for various BH masses and spins.
As in the previous work \cite{Cho15b}, we assume the mass range of the fiducial NS  as $1 \leq \Mns/\msun \leq 2$,
and the BH mass range as $4 \leq \Mbh/\msun \leq 10$. We only consider a single detector and the SNR is assumed to be 10.
On the other hand, while we considered a nonspinning system hence only the two mass parameters were used  in the previous work,
we consider an aligned-spin system here, and the BH spin parameter is also taken into account  in this work.

\begin{figure}[t]
\begin{center}
\includegraphics[width=8.0cm]{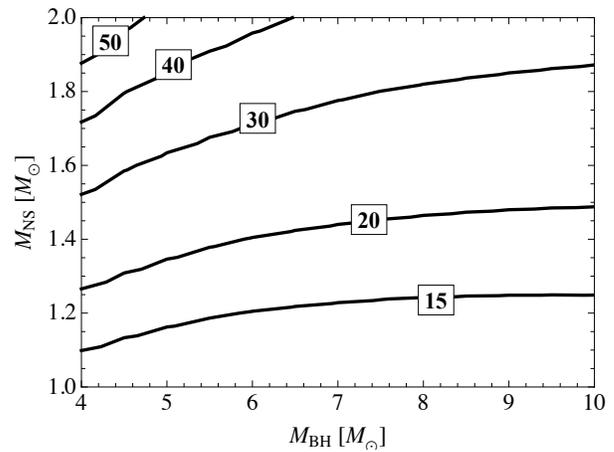}
\end{center}
\caption{Contours of constant measurement errors  ($\sigma/M_{\rm NS} \times 100$) in parameter estimation
for  BH-NS binaries. We assumed the BH spin of $\chi_{\rm BH}=0$ and the SNR of $\rho=10$.}\label{fig.1}
\end{figure}

In Fig.~\ref{fig.1}, we show the measurement errors in $\Mns$  in the $\Mbh - \Mns$ plane, where the values indicate the fractional errors ($\sigma / \Mns)$ in percentage.
Here, the BH spin is assumed to be 0. 
We can find that the fractional error is the largest with the heaviest NS and the lightest BH binary (upper left corner), and tends to decrease with increasing BH mass or decreasing NS mass.
This trend is similar to the result for the nonspinning system given in \cite{Cho15b}.
However, the measurement accuracy is significantly decreased compared to that of the nonspinning system.
The  errors in Fig.~\ref{fig.1} are in the range of about $10\% - 50 \%$, and those are overall about 20 times larger than the errors for the nonspinning system given in \cite{Cho15b}.
The main cause of this difference is that the parameter space to be explored is extended from a four parameter system to a five parameter one  by taking into account the spin parameter additionally.
In general, if a binary system has spins, a degeneracy between the components' mass ratio and their spins significantly degrades our ability to
measure the individual component masses \cite{Cut94,Poi95}.
Therefore, in the posterior distribution of parameter estimation for an aligned-spin system, the mass parameters are strongly correlated with
the spin parameter; thus, the range of error in $\Mns$ can be significantly increased compared to that for a nonspinning system (for example, see \cite{Osh14,Cho15d}).

\begin{figure}[t]
\includegraphics[width=8.0cm]{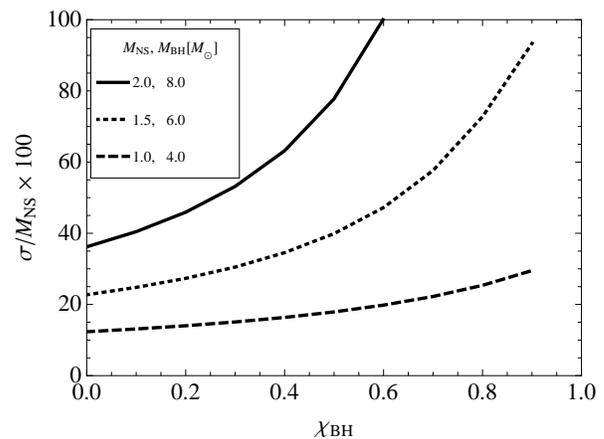}
\caption{Measurement errors  ($\sigma/M_{\rm NS} \times 100$) calculated by changing the BH spin ($\chi_{\rm BH}$). We assume a fixed mass ratio as $M_{\rm BH}/M_{\rm NS}=4$ and the SNR of $\rho=10$.}\label{fig.2}
\end{figure}

In Fig.~\ref{fig.2}, we show a dependence of the fractional errors on the BH spin.
Here, we calculate the errors only by varying  the BH spin from 0 to 0.9 for three NS masses of $2~\msun$, $1.5~\msun$ and $1~\msun$
assuming a fixed mass ratio of $M_{\rm BH}/M_{\rm NS}=4$.
Generally, the spin-mass correlation tends to be stronger as the BH spin increases; thus, a high BH spin lowers the measurement accuracy of the masses.
We therefore can find that the errors become larger as the BH spin  increases.
We can also find that the increasing tendency is more pronounced for higher NS masses (or higher total masses).
The fractional errors can increase  from $23 \%$ to $93 \%$ and  from $12 \%$ to $30 \%$ for $\Mns=1.5~\msun$ and $1~\msun$, respectively.
Especially, for the case of $\Mns=2~\msun$, we found that the fractional errors can increase from $36 \%$ to $182 \%$, and
those can be larger than the true value of $\Mns$ if $\chibh>0.6$.
Most interestingly, in certain cases we cannot distinguish between BHs and NSs.
For example, if the true values of a GW signal are $\Mns, \Mbh=2, 8~\msun$ and $\chibh=0.9$, the recovered NS mass lies in about $\Mns \lesssim 5.6~\msun$,
then we cannot determine whether this is a NS or a light BH (for more examples, see \cite{Han13,Lit15,Man15}).


\section{Discussion}

In this work, we investigated the measurement accuracy of the NS mass in parameter estimation of GWs from BH-NS binaries.
We adopted the 3.5pN aligned-spin TaylorF2 waveform model in which the spin terms are included up to 2.5pN,  
and applied this  model to the FM method.
In our BH-NS binaries, we assumed the BH mass to be lower than $10~\msun$ and the spin to be aligned with the orbital angular momentum.
The NS mass is assumed to be $1\leq  \Mns/\msun \leq 2$ and the NS spin is not considered in the analysis.
The result shows that the fractional errors ($\sigma/M_{\rm NS} \times 100$) are in the range of $10\% - 50 \%$  in our mass region for a given BH spin $\chibh=0$,
and these errors become larger as the BH spin increases.
In particular,  by comparing our result with our previous work \cite{Cho15b} where we considered a nonspinning system,
we confirmed that  the mass ratio is strongly correlated with the spin in an aligned-spin system.
The errors are overall about 20 times larger than those for the nonspinning system in the same mass region.

Typically, the evolution of
merging binaries containing BHs is divided into three phases;
inspiral, merger and ringdown.
In the low mass region as considered in this work the merger-ringdown phases are placed out of the detector
sensitivity band; thus, only the inspiral waveform models such as TaylorF2 can be used in the GW data analysis.
However, for more massive systems,
full inspiral-merger-ringdown waveform models should be used not to lose the merger-ringdown portions.
For this purpose, various models have been developed over the past years,
and the phenomenological models have been commonly used \cite{Aji07,Aji08,Aji08b,Aji11b,San10,Han14,Kha16} (for a brief description of these models, see \cite{Cho15c}).
Since the phenomenological models are constructed as analytic functions in the frequency-domain,
those are also applicable  to the FM method.
Therefore, our work can be easily extended to higher mass BBHs.

Finally, it is worth noting that the orbital plane can precess if the spin is misaligned with the orbital angular momentum.
For  space-based detectors such as LISA, several works showed that the precession effect can break the mass-spin degeneracy;
thus, improve the measurement accuracy of the mass and spin parameters \cite{Lan06,Kle09}.
The authors in \cite{Cho13,Osh14b,Cat14} showed  that such improvement can also be achieved for ground-based detectors.
One of the phenomenological models, i.e. PhenomP \cite{Han14}, was designed to model the precessing BBH waveforms;
thus, using this model our analysis can also be extended to a precessing system.

\begin{acknowledgments}
This work was supported by the National Research Foundation of Korea (NRF) grant funded by the Korea government (Ministry of Science, ICT \& Future Planning) (No. 2016R1C1B2010064).
This work used the computing resources at the KISTI Global Science Experimental Data Hub Center (GSDC).
\end{acknowledgments}

\end{document}